\documentclass[11pt]{article}
\pdfoutput=1
\usepackage{graphics}
\usepackage{hyperref}
\usepackage{graphicx}
\usepackage{epsfig}
\usepackage{epsf}
\usepackage{epstopdf}
\usepackage{amsmath}  
\usepackage{ tipa }
\usepackage{braket}
\usepackage{xcolor}

\def\bear{\begin{eqnarray}}
\def\ear{\end{eqnarray}\noindent}
\newcommand{\no}{\noindent}

\title{\boldmath One-loop Amplitudes in the Worldline formalism.}

\author{James P. EDWARDS~$^\dag$ , \underline{C. Moctezuma MATA}~$^\dag$   and Christian SCHUBERT~$^\dag$ 
\\
$^\dag$~Intituto de F\'isica y Matem\'aticas, Universidad Michoacana de San Nicol\'as de Hidalgo\\
Apdo. Postal 2-82, C.P. 58040, Morelia, Michoacan, Mexico
}

\begin{document}
\maketitle
\flushbottom

\begin{abstract}
We summarize recent progress in applying the worldline formalism to the analytic calculation of one-loop $N$-point amplitudes.
This string-inspired approach is well-adapted to avoiding some of the calculational inefficiencies of the standard Feynman diagram approach, 
most notably by providing master formulas that sum over diagrams differing only by the position of external legs and/or internal propagators. 
We illustrate the mathematical challenge involved with the low-energy limit of the $N$-photon amplitudes in scalar and spinor QED,
and then present an algorithm that, in principle, solves this problem for the much more difficult case of the $N$-point amplitudes at full momentum in
$\phi^3$ theory. The method is based on the algebra of inverse derivatives
in the Hilbert space of periodic functions
orthogonal to the constant ones, in which the Bernoulli numbers and polynomials play a central role. 
\end{abstract}

\section{Introduction.}
The worldline formalism of quantum field theory is based on first-quantized relativistic particle path integrals and  provides an alternative to Feynman diagrams in the construction of the perturbation series in QFT. Introduced by Feynman in 1950/1 for QED \cite{feynman50,feynman51}, but then largely forgotten, this formalism began to gain popularity after the work of Bern and Kosower who managed to simplify the calculation of scattering amplitudes in quantum chromodynamics \cite{BernKosower}  using the fact that string theory reduces to quantum field theory in the limit where the string tension becomes infinite. Moreover, along these lines they established a set of rules which allows one to construct the 
one-loop gluon amplitudes at the parameter integral level without referring to string theory any more. 
\\
After this had made it clear that techniques from first-quantised string perturbation theory have the potential to improve on the efficiency of field theory calculations, this was further investigated by Strassler, who derived the Bern-Kosower-type rules for one-loop effective actions of scalars, Dirac spinors, and vector bosons in a background gauge field  within the framework of QFT without the use of either string theory or Feynman diagrams.
\\
During the last three decades the worldline formalism has been applied to a steadily expanding circle of problems in QFT, providing new computational options as well as useful physical intuition (for reviews, see \cite{ChrisRev,ChrisAndJames}). 
\\
Here we report on recent work that focuses on a particular feature of this formalism
which is to generate master formulas that effectively sum up Feynman diagrams differing only by
the position of the external legs and/or of internal propagator insertions. While this property has
already played an important role in a number of applications based on numerical
calculation or semiclassical approximation, either at the path-integral or the parameter-integral
level, at the moment it remains a formidable mathematical challenge to develop techniques that
would allow one to perform this type of master integral analytically without breaking it up into
the sectors that would - usually up to some integration by parts (`IBP')  - correspond to individual
Feynman diagrams. 
\\
We will show here how to do this in principle for the simplest but fundamental case of abelian
one-loop amplitudes. 
After briefly recalling some basic concepts of the formalism we present results for the low-energy limit of the N-photon amplitude both for scalar and spinor QED, obtained in a way that avoids the necessity to sum over ``crossed'' diagrams. 
Subsequently, we discuss the natural  appearance of the Bernoulli numbers and polynomials in the worldline formalism, and then outline a new method to compute the general scalar $N$-point integral. 
\\
\section{The worldline formalism in quantum field theory.}

The starting point for deriving the path integral representation of the $x$-space matrix element is the Green's function for the covariantized Klein-Gordon operator  $(\partial + ie A)^2 + m^2$,
\bear
D^{xx'}[A]
 \equiv
\big\langle x' \big| {1\over -(\partial + ie A)^2 + m^2} \big| x \big\rangle\,.
\ear
We use natural units $\hbar = c=1$ and work throughout in Euclidean space.
\\
The next step is to introduce a Schwinger proper-time parameter $T$ in order to exponentiate the denominator:

\bear
D^{xx'} [A] &=& 
\Big\langle x' \Big| \int_0^{\infty}dT\, {\rm exp}\Bigl\lbrack - T (  -(\partial + ie A)^2+m^2)\Bigr\rbrack  \Big| x \Big\rangle\,.
\ear
%
%
%
%
%
%
%

Expressing the matrix element in its path integral representation we get, after some manipulations,

\bear
D^{xx'} [A] &=& 
\int_0^{\infty}
dT\,
e^{-m^2T}
\int_{x(0)=x}^{x(T)=x'}
{\cal D}x(\tau)\,
e^{-\int_0^T d\tau \bigl(\frac{1}{4}\dot{x}^2 +ie\dot x\cdot A(x(\tau))\bigr)} \,.
\label{scalpropfreepi}
\ear\no
\no

We aim to study photon amplitudes so we fix the background field as a sum of $N$ plane waves representing their asymptotic states,

\bear
A^{\mu}(x(\tau)) = \sum_{i=1}^N \,\varepsilon_i^{\mu} e^{ik_i\cdot x(\tau)} \,.
\label{pw}
\ear

\begin{figure}[htbp]
\begin{center}
 \includegraphics[width=0.65\textwidth]{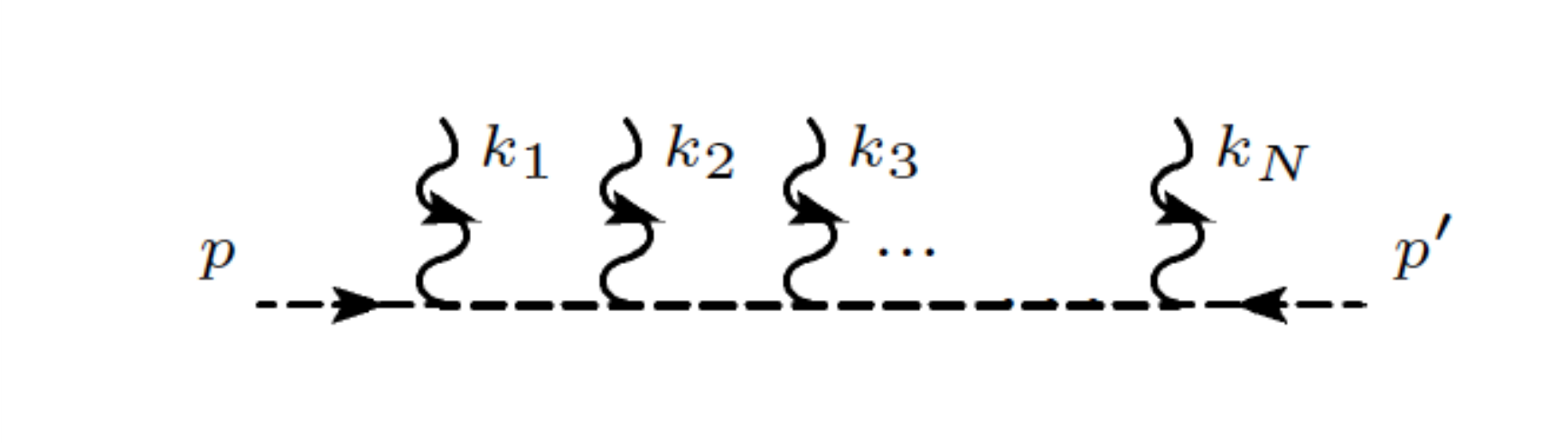}
\caption{Scalar propagator dressed with $N$ photons.}
\label{fig-propexpand}
\end{center}
\end{figure}

The photon-dressed propagator shown in Fig. \ref{fig-propexpand} can be obtained by Fourier transforming the endpoints. We stress that summation over the $N!$ permutations of the photons along the line is understood.

Analogously it can be shown that the path integral representation for the one-loop effective action is:
\bear
\Gamma_{\rm scal} [A] &=&- {\rm Tr}\,{\rm ln} \Bigl\lbrack -(\partial + ie A)^2+m^2\Bigr\rbrack  
= \int_0^{\infty} \frac{dT}{T} \,{\rm Tr} \, {\rm exp}\Bigl\lbrack - T (  -(\partial + ie A)^2+m^2)\Bigr\rbrack  \nonumber\\
&=&
\int_0^{\infty}
\frac{dT}{T}\,
e^{-m^2T}
\int_{x(0)=x(T)}
{\cal D}x(\tau)\,
e^{-\int_0^T d\tau \bigl(\frac{1}{4}\dot{x}^2 +ie\dot x\cdot A(x(\tau))\bigr)}\, .
\label{EAscal}
\ear

To extract the one-loop N-photon amplitude  we choose the plane-wave background \eqref{pw} and an expansion to $N$th order in the coupling:
\bear
\hspace{-1em}\Gamma_{\rm scal}[\lbrace k_i,\varepsilon_i\rbrace]
=
(-ie)^{N}
\int \frac{dT}{T} e^{-m^2T}
\int_{x(0)=x(T)} {\cal D}x\, e^{-\int_0^T d\tau \frac{1}{4}\dot{x}^2}\,
V_{\rm scal}^\gamma[k_1,\!\varepsilon_1]\ldots
V_{\rm scal}^\gamma[k_N,\!\varepsilon_N]
\, .
\label{Nvertop}
\ear\no

\begin{figure}[htbp]
\begin{center}
\includegraphics[width=0.45\textwidth]{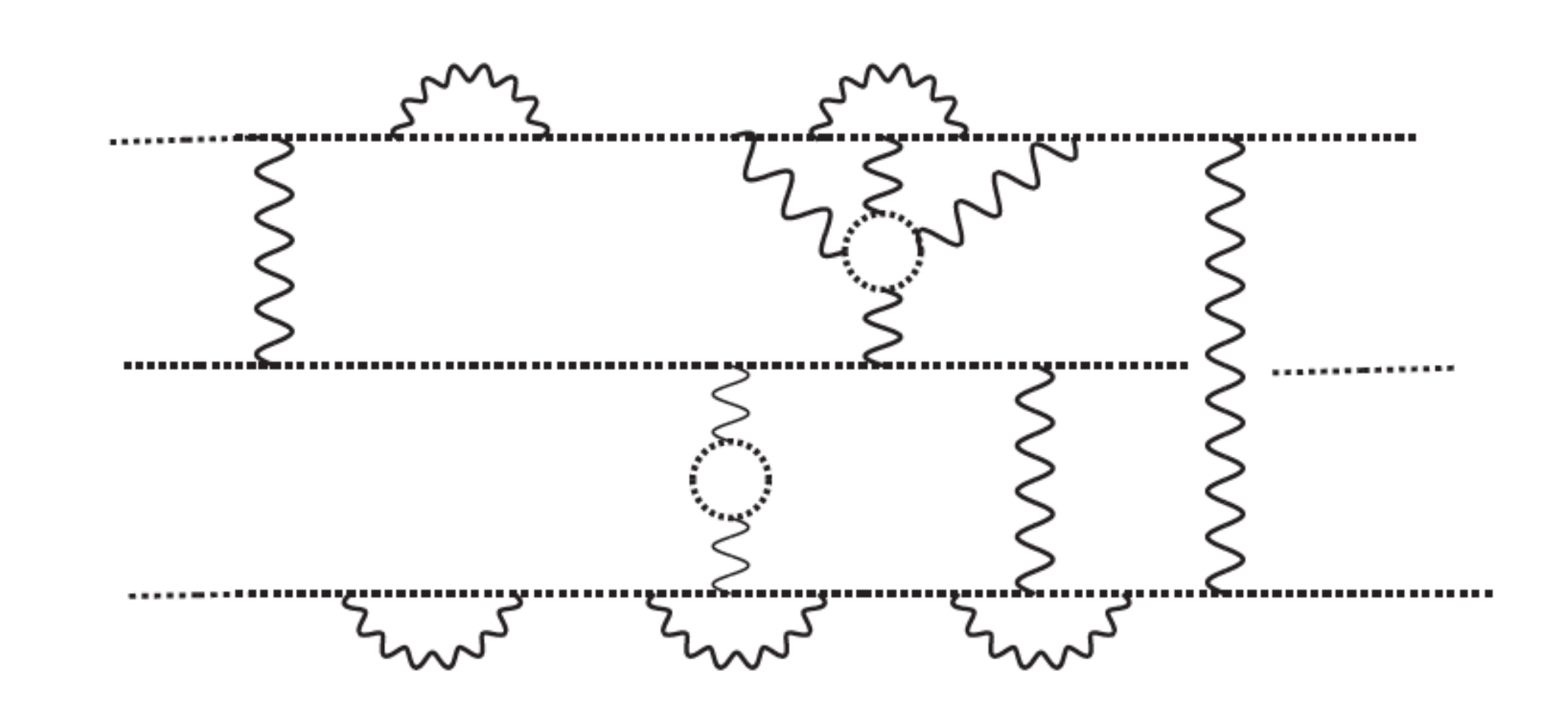}
\caption{A typical multiloop diagram in scalar QED.}
\label{fig-QEDSmatrix}
\end{center}
\end{figure}

In the last expression  $V_{\rm scal}^\gamma$ denotes the photon
vertex operator (note that this vertex is the same as the one used in string perturbation theory),
\bear
V_{\rm scal}^\gamma[k,\varepsilon]
=
\int_0^Td\tau\,
\varepsilon\cdot \dot x(\tau)
\,{\rm e}^{ik \cdot x(\tau)}
\label{defvertopscal}
\ear
The formulas above are valid off-shell so one can obtain multi-loop amplitudes from the previous results by sewing pairs of photons. See, e.g.,
Fig. \ref{fig-QEDSmatrix}. 

\begin{figure}[htp]
\hspace{15.5pt}
\centerline{\includegraphics[width=13cm]{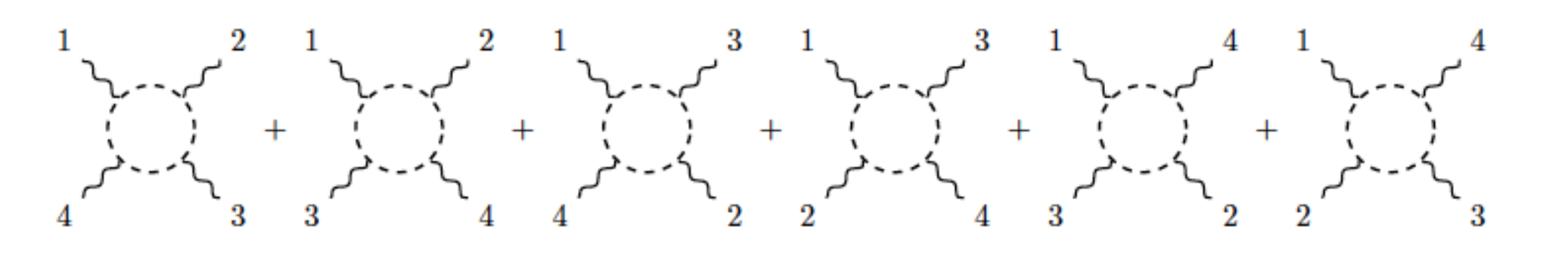}}
\caption{Feyman diagrams for photon-photon scattering.}
\label{4-photon}
\end{figure}

In our perturbative treatment, one first has to fix the zero mode that comes from the translation invariance of the free path integral. 
It is then a matter of standard Wick contraction combinatorics
\cite{ChrisRev} to obtain from equation \eqref{Nvertop}
the following master formula for the scalar QED one-loop N-photon amplitude,
valid both on- and off-shell:
\bear
\Gamma[\lbrace k_i,\varepsilon_i\rbrace]
&=&(-ie)^n (2 \pi)^D \delta \left( \sum k_i \right) \int_0^\infty \frac{dT}{T} (4 \pi T)^{-D/2}e^{- m^2 T} \prod_{i=1}^n \int_0^T d\tau_i
\nonumber
\\
&&\times{\rm exp} \Big\{ \sum_{i,j=1}^n \Big[ \frac{1}{2} G_{Bij} k_i \cdot k_j - i \dot{G}_{Bij} \varepsilon_i \cdot k_j + \frac{1}{2} \ddot{G}_{Bij} \varepsilon_i \cdot \varepsilon_j\Big]\Big\}\Big|_{\varepsilon_1 \, \, \varepsilon_2 \dots \varepsilon_n}\,. \nonumber\\
\label{master}
\ear
Here is understood that only the terms linear in all the $\varepsilon_1  \dots \varepsilon_n$ have to be taken.
$G_{Bij}$ is the bosonic Green's function of the operator $-\frac{1}{4} \frac{d^2}{d\tau^2}$ with periodic boundary conditions  in the space orthogonal to the zero mode and $\dot{G}_{Bij}$ , $\ddot{G}_{Bij}$ are its first and second derivatives with respect to $\tau_i$,
\bear
G_{Bij} &=& |\tau_i - \tau_j| - \frac{(\tau_i-\tau_j)^2}{T},
\\
\dot{G}_{Bij} &=& {\rm sign}(\tau_i-\tau_j)-2 \frac{(\tau_i-\tau_j)}{T},
\\
\ddot{G}_{Bij} &=& 2 \delta(\tau_i-\tau_j)-\frac{2}{T}.
\ear
For our present purpose, it is important to note that the usual summation over inequivalent crossed diagrams is implicit in the master formula in the integration over all possible orderings of the insertion points of the photon legs (the $\tau_{i}$) around the loop. 
For example, in the four-photon case the master integral consists of 6 sectors, which in a standard calculation would correspond to the six diagrams shown in Figure \ref{4-photon}
(in spinor QED; for scalar QED, there are additional diagrams involving the seagull vertices). 

Another advantage of the formalism is that the relation between scalar and spinor QED is more evident  as usual; namely from \eqref{EAscal} we can,
up to the global normalization, 
obtain the path integral representation of the effective action for spinor QED by the insertion of a {\it spin factor} ${\rm Spin}[x,A]$, 
\bear
{\rm Spin}[x,A] = {\rm tr}_{\Gamma} {\cal P}
\exp\Biggl[{{i\over 4}e\,[\gamma^{\mu},\gamma^{\nu}]
\int_0^Td\tau F_{\mu\nu}(x(\tau))}\Biggr] \, .
\label{spinfactor}
\ear
The notation $\cal P$ means path-ordering, which is necessary in view of the fact that the exponents at different proper times will not commute so the exponential function in general will not be an
ordinary one. Alternatively one may use the ``cycle replacement rule''
\cite{BernKosower,Strassler}
 to convert the parameter integrals for the amplitudes in scalar QED into their corresponding expressions for spinor QED.

\section{Low-energy limit of the N - photon amplitudes.}

The condition that defines the low-energy limit at the one-loop level is that all photon energies be small compared to the scalar mass:
\bear
\omega_i \ll m , \, \, \, \, i=1, \dots N.
\ear
It is easy to show that in the worldline formalism this limit can be implemented by 
truncating the photon vertex operators to their terms linear in the momentum:
\bear
V_{scal}^{(\mathrm{LE})}[k,\varepsilon]&=& \int_0^T d\tau \, \varepsilon \cdot \dot{x} (\tau) i k \cdot x(\tau)\,.
\ear
By adding a total derivative, it is possible to rewrite this vertex operator in terms of the photon field strength tensor 
$f^{\mu \nu}\equiv   k^\mu\varepsilon^\nu-\varepsilon^\mu k^\nu $,
\bear
\hspace{-1em}V_{scal}^{(\mathrm{LE})}[f]&=&V_{scal}^{(\mathrm{LE})}[k,\varepsilon]-\frac{i}{2} \int_0^T d\tau \frac{d}{d\tau}(\varepsilon \cdot x(\tau) k \cdot x(\tau))
\nonumber
\\
&=&\frac{i}{2} \int_0^T d\tau x(\tau) \cdot f \cdot \dot{x}(\tau).
\ear
The Wick contraction of a product of the vertices will produce terms that by suitable IBP can be written as integrals of ``$\tau$-cycles'', which are products of $\dot{G}_{B_{ij}}$s with the indices forming a closed chain, $\dot{G}_{B_{i_1 i_2}} \dot{G}_{B_{i_2 i_3}} \cdots \dot{G}_{B_{i_n i_1}}$. 
And these $\tau$-cycles will appear as ``bicycles", namely together with a corresponding field-strength factor:
\bear
\int_0^T d\tau_{i_1} \dots \int_0^T d\tau_{i_n} \dot{G}_{i_1 i_2} \dot{G}_{i_2 i_3} \cdots \dot{G}_{i_n i_1}  {\rm tr}(f_{i_1} f_{i_2} \cdots f_{i_n} )\,.
\ear
In the one-dimensional worldline QFT, we can identify such a bicycle with
the one-loop n-point Feynman diagram of Fig. \ref{npointLow}:

\begin{figure}[htbp]
\begin{center}
 \includegraphics[width=0.15\textwidth]{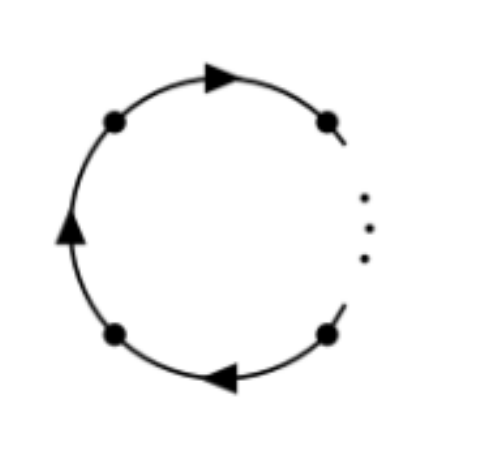}
\caption{Worldline Feynman diagram representing an integrated bicycle factor.}
\label{npointLow}
\end{center}
\end{figure}

It will thus be useful to introduce the "Lorentz cycles'' $Z_n$,
\bear
Z_2(ij) &\equiv& \varepsilon_i \cdot k_j \varepsilon_j \cdot k_i - \varepsilon_i \cdot \varepsilon_j k_i \cdot k_j=\frac{1}{2} \mathrm{tr}(f_i f_j),
\\
Z_n(i_1 i_2 \dots i_n) &\equiv& {\rm tr}(f_{i_1} f_{i_2} \cdots f_{i_n} ), \quad n \geq 3.
\ear

By simple combinatorics, we arrive at:
\bear
\hspace{-1em}\left<  V_{\rm scal}^{\gamma ({\rm LE})} [ f_1] \dots V_{\rm scal}^{\gamma ({\rm LE})} [ f_n]  \right> = i^n T^n {\rm exp} \left\{  \sum_{m=1}^\infty b_{2m} \sum_{\{i_1 \dots i_{2m} \} } Z_{2m}^{\rm dist }(\{i_1 i_2  \dots i_{2m} \})\right\}\Bigg|_{f_1 . . . f_N},
\nonumber\\
\ear 
where $ Z_{2m}^{\rm dist }(\{i_1 i_2  \dots i_{2m} \})$ denotes the sum over all distinct Lorentz cycles  which can be formed with a given subset of indices, and $b_n$ denotes the basic ``bosonic cycle integral"
\bear
b_n \equiv \int_0^1 du_1 du_2 \dots du_n \dot{G}_{B12} \dot{G}_{B23} \cdots \dot{G}_{Bn1}\, .
\label{chain}
\ear
It can be shown that this integral can be expressed in terms of the Bernoulli numbers $\mathcal{B}_n$
as \cite{5}
\bear
b_n = \Big\{ \begin
{matrix}
-2^n \frac{\mathcal{B}_n}{n!} & n & {\rm even}\, ,
\\
0 & n & {\rm odd}\, .
\end{matrix}
\ear
Introducing further $f_{\rm tot} \equiv \sum_{i=1}^nf_i$, and using the combinatorial fact that 
\bear
{\rm tr}[ (f_1+ \dots + f_N)^n]|_{\rm all\, different} = 2n  \sum_{\{i_1 \dots i_{n} \} } Z_{n}^{\rm dist }(\{i_1 i_2  \dots i_{n} \}),
\ear
from \eqref{Nvertop} we obtain the following formula for the low-energy limit of the one-loop N-photon amplitude (now with $D=4$):
\begin{eqnarray}
\Gamma_{\mathrm{scal}}^{\mathrm(LE)}(k_1,\varepsilon_1 ; . . . ;k_N, \varepsilon_N) = \frac{e^N \Gamma(N-2)}{(4 \pi)^2 m^{2N-4}}\mathrm{exp} \left\{ \sum_{m=1}^\infty \frac{b_{2m}}{4m} \mathrm{tr} (f_{\mathrm{tot}}^{2m})     \right\}\Bigg|_{f_1 . . . f_N}.
\end{eqnarray}
This result can be further simplified if we consider the case where all photons have helicity '+'. It turns out that in this case \cite{56}
\bear
f_{\rm tot}^2 = - \chi_+ 
\ear
with
\bear
\chi_+ \equiv \frac{1}{2} \sum_{1 \leq i < j \leq N } [k_i k_j]^2 \, .
\ear
In this special case:
\bear
{\rm tr} (f_{\rm tot}^{2m}) = 4 (-1)^m (\chi_+)^m \, .
\ear
We can therefore rewrite 
\bear
{\rm exp} \Bigg\{ \sum_{m=1}^\infty \frac{b_{2m}}{4m} {\rm tr}(f_{\rm tot}^{2m})  \Bigg\}\Bigg|_{f_1 \dots f_N} = {\rm exp} \Bigg\{ \sum_{m=1}^\infty (-1)^m \frac{b_{2m}}{m} \chi_+^m \Bigg\}\Bigg|_{\rm all \, \,  different}.
\ear
Note that using the series expansion
\bear
\frac{\chi_+}{{\rm sin}^2 \sqrt{\chi_+}} = - \sum_{k=0}^\infty (-1)^k \frac{2k-1}{(2k)!}2^{2k} \mathcal{B}_{2k} \chi_+^k,
\ear
we can recognize the exponent as the series expansion of $-2 {\rm ln} \frac{{\rm sin} \sqrt{\chi_+}}{\sqrt{\chi_+}}$, and finally obtain for the low-energy limit of the N-photon ``all +'' amplitudes \cite{51} 
\begin{eqnarray}
\Gamma_{\mathrm{scal}}^{(\mathrm{LE})}(k_1,\varepsilon_1 ; . . . ;k_N, \varepsilon_N) = -\frac{m^4}{(4 \pi)^2} \left( \frac{2 i e}{m^2}\right)^N \frac{\mathcal{B}_N}{N(N-2)}\chi_N^+ 
\end{eqnarray}
where we have further defined
\begin{eqnarray}
\hspace{-1em}\chi_N^+ \equiv \chi_+^{N/2}|_{\mathrm{all\, different}} = \frac{(N/2)!}{2^{N/2}} \left\{ [k_1 k_2]^2 [k_3 k_4]^2 \cdots [k_{N-1} k_N]^2 + \mathrm{all \, \, distinct \, \, \, permutations}\right\}. 
\nonumber\\
\end{eqnarray}
As a consequence of the above-mentioned cycle replacement rule, 
the transition from scalar to spinor QED can be done simply by changing the chain integral \eqref{chain} to the ``super chain integral'' \cite{BernKosower,Strassler,ChrisRev,5}
\begin{eqnarray}
\int_0^1du_1 . . . du_n (\dot{G}_{B12}\dot{G}_{B23} \cdots \dot{G}_{Bn1} 
- G_{F_{12}} G_{F_{23}} \cdots G_{F_{n 1}}) = (2-2^n)b_n.
\label{superchain}
\end{eqnarray}
The only other change is a global factor of (-2) for statistics and degrees of freedom, consequently:
\begin{eqnarray}
\Gamma_{\mathrm{spin}}^{\mathrm(LE)}(k_1,\varepsilon_1 ; . . . ;k_N, \varepsilon_N) =-2 \frac{e^N \Gamma(N-2)}{(4 \pi)^2 m^{2N-D}}\mathrm{exp} \left\{ \sum_{m=1}^\infty (1-2^{2m-1})\frac{b_{2m}}{2m} \mathrm{tr} (f_{\mathrm{tot}}^{2m})     \right\}\Bigg|_{f_1 . . . f_N}.
\nonumber\\
\end{eqnarray}
We have started with this particular calculation because
the integrals \eqref{chain} resp. \eqref{superchain} display in a nutshell the basic issue that we wish to address here. Although the master formula \eqref{master} contains all possible orderings
of the photon legs around the loop, this advantage appears to get lost when it comes to the
actual computation of these integrals, due to the sign function contained in $\dot G_B$ that 
at first sight appears to make it necessary to split the integrals into ordered sectors. 
Doing so would lead to integrals that are individually trivial, but arriving at the displayed closed-form results
in terms of the Bernoulli numbers in this way would be quite hard (the intrepid reader may wish to try!).

\section{Bernoulli numbers and polynomials in the worldline formalism.}

As we have seen in the previous section, the use of this formalism to obtain one-loop photon amplitudes in scalar QED in the low-energy limit involves exclusively the computation of bosonic cycle integrals which depend on the first derivative of the bosonic Green function. 
For general momenta, expansion of the master formula \eqref{master} yields
an integrand with a certain polynomial $P_N$ depending on the first and second derivatives of this Green function as well as on the kinematic invariants. 
\bear
{\rm exp} \Big\{ \Big\}\Big|_{\varepsilon_1 \dots \varepsilon_N} = (- i)^N P_N(\dot{G}_{Bij},\ddot{G}_{Bij}){\rm exp} \Big[ \frac{1}{2}  \sum_{i,j=1}^N G_{Bij} k_i \cdot k_j\Big].
\label{expand}
\ear
By suitable partial integrations in the variables $\tau_1, \dots, \tau_N$ we can remove all the second derivatives $\ddot{G}_{Bij}$:
\bear
P_N(\dot{G}_{Bij},\ddot{G}_{Bij}) {\rm e}^{ \frac{1}{2}  \sum_{i,j=1}^N G_{Bij} k_i \cdot k_j} \longrightarrow Q_N(\dot{G}_{Bij}){\rm e}^{ \frac{1}{2}  \sum_{i,j=1}^N G_{Bij} k_i \cdot k_j}.
\label{PolyExp}
\ear
Now in addition to the sign functions contained in the $\dot G_B$s we have to deal with the
absolute value functions appearing in the $G_B$ in the exponential. Thus if we wish to avoid 
a break-up into ordered sectors we are confronted with a very non-standard integration problem.
However, progress can be made using a basis of inverse derivatives. 
Recall that we are working in the Hilbert space  of periodic functions orthogonal to the constant functions. 
In this space the ordinary nth derivative is invertible and the integral kernel of the inverse is essentially given by the nth Bernoulli polynomial \cite{15}
\bear
\bra{u_i} \partial^{-n} \ket{u_j} = -\frac{1}{n!}B_{n}(|u_i-u_j|){\rm sign}^n(|u_i-u_j|).
\label{Inverse}
\ear
Note that \eqref{Inverse} reduces to the Bernoulli numbers in the case $u_i=u_j$,
\bear
\bra{u_i}\partial^{-2n} \ket{u_i}=-\frac{B_n}{n!} \equiv - \hat{B}_n.
\label{BernoulliN}
\ear
In terms of \eqref{Inverse} and \eqref{BernoulliN} we can expand each exponential in \eqref{expand}
using the identity \cite{135}
\bear
{\rm e}^{k_{ij} G_{Bij}}= 1 + 2 \sum_{n=1}^\infty k_{ij}^{n-1/2} {\rm H}_{2n-1} \left( \frac{\sqrt{k_{ij}}}{2} \right) \overline{\bra{u_i}\partial^{-2n} \ket{u_j}}.
\label{magic}
\ear
Here the ${\rm H}_n(x)$ are Hermite polynomials, and we have abbreviated
\bear
\overline{\bra{u_i}\partial^{-2n} \ket{u_j}} \equiv \bra{u_i}\partial^{-2n} \ket{u_j}-\bra{u_i}\partial^{-2n} \ket{u_i}.
\ear
We rescale the $\tau_i$ integrals to the unit circle, $\tau_i= T u_i$. 

To show the potential of 
identity \eqref{magic} to solve the problem of circular integration, let us restrict ourselves here to
the simplest case of the one-loop N-point amplitude for $\phi^3$-theory in D dimensions. 
Here the master formula corresponding to \eqref{master} is \cite{ChrisRev}
\bear
I_N (p_1, \dots p_N) &=& \frac{1}{2} (4 \pi)^{-d/2} (2 \pi)^d \delta \left( \sum_{i=1}^N p_i \right) \hat{I}_N(p_1, \dots , p_N ),
\\
 \hat{I}_N(p_1, \dots , p_N ) &=& \int_0^\infty \frac{dT}{T} T^{N-d/2} e^{-m^2 T} \int_0^1 du_1 \cdots \int_0^1 du_N e^{T \sum_{i<j=1}^N G_{Bij} p_i \cdot p_j }.
 \nonumber
 \\
\ear
Let us have a look at the three-point case. Applying \eqref{magic} yields
\begin{align}
{\rm e}^{ G_{B12} p_{12} +G_{B13} p_{13}+ G_{B23} p_{2 3}} &=\Big\{ 1 + 2 \sum_{i=1}^\infty ( p_{12})^{i-1/2} {\rm H}_{2i-1} \left( \frac{\sqrt{  p_{12}}}{2} \right) \big[\bra{u_1}\partial^{-2i} \ket{u_2}+\hat{B}_{2i}\big] \Big\}
\nonumber
\\
 &\times \Big\{ 1 + 2 \sum_{j=1}^\infty ( p_{13})^{j-1/2} {\rm H}_{2j-1} \left( \frac{\sqrt{ p_{13}}}{2} \right) \big[\bra{u_1}\partial^{-2j} \ket{u_3}+\hat{B}_{2j}\big] \Big\}
\nonumber
\\
 &\times \Big\{ 1 + 2 \sum_{k=1}^\infty (p_{23})^{k-1/2} {\rm H}_{2k-1} \left( \frac{\sqrt{  p_{23}}}{2} \right) \big[\bra{u_2}\partial^{-2k} \ket{u_3}+\hat{B}_{2k}\big] \Big\}.
\nonumber
\\
\label{3point}
\end{align}
where we abbreviated $p_{ij}=T p_i \cdot p_j $. 
When we integrate eq. \eqref{3point}, the three $\bra{u_i} \partial^{-2n} \ket{u_j}$  
individually cannot
contribute because $\int_0^1 du_i \bra{u_i} \partial^{-2n} \ket{u_j}=\int_0^1 du_j \bra{u_i} \partial^{-2n} \ket{u_j}=0$, and the only non-trivial integration can be done by applying the completeness relation $\int_0^1 du \ket{u}\bra{u} = \mathrm{1}$. In this way, we obtain
\bear
\int du_1 du_2 du_3 \bra{u_1} \partial^{-2i} \ket{u_2} \bra{u_1} \partial^{-2j} \ket{u_3} \bra{u_2} \partial^{-2k} \ket{u_3} = -\hat{B}_{2(i+j+k)} \, ,
\ear
and get a closed form-expression for the coefficients
\bear
\hspace{-2em}\int du_1 du_2 du_3 G_{B12}^a G_{B13}^b G_{B23}^c = a! b! c! \sum_{i=\lfloor 1+a/2 \rfloor}^a \sum_{j=\lfloor 1+b/2 \rfloor}^b
\sum_{k=\lfloor 1+c/2 \rfloor}^c h_i^a h_j^b h_k^c (\hat{B}_{2i}\hat{B}_{2j}\hat{B}_{2k}-\hat{B}_{2(i+j+k)})\, .
\nonumber
\\
\ear
Here we have assumed that a,b,c are all different from zero. The coefficients $h_i^a$ are given by
\bear
h_i^a = (-1)^{a+1} \frac{2(2i-1)!}{(2i-a-1)!(2a-2i+1)!}\, ,
\ear
which one can read off from
\bear
H_n(x) = n! \sum_{m=0}^{\lfloor \frac{n}{2} \rfloor} (-1)^m \frac{(2x)^{n-2m}}{m!(n-2m)!}\, .
\ear
Starting from the four-point case, the use of \eqref{magic} does not immediately lead to integrals that can all be done just by applying the completeness relation
(such integrals are called ``chain integrals''), but it can be shown that it is always possible to achieve a complete reduction to chain integrals
using an integration-by-parts algorithm that was initially developed for a somewhat different purpose \cite{34}.

\section{Conclusion.}

To summarize, we have presented here a new approach to the mathematical challenge of
the analytic calculation of one-loop $N$-point amplitudes using worldline master integrals
without breaking them up into the sectors corresponding to individual Feynman diagrams. 
For scalar integrals, the problem has been
solved in principle, but it remains to establish the correspondence between the resulting multiple sums 
with the hypergeometric functions that are known to describe these amplitudes. 
The generalization to the case of the QED $N$-photon amplitudes is in progress. 

\pdfbookmark[1]{References}{ref}

\end{document}